\documentclass{PoS}

\PoS{PoS(HEP2005)336}

\title{Charged Higgs production: higher-order corrections}

\ShortTitle{Charged Higgs production}

\author{\speaker{Nikolaos Kidonakis}\\
       Kennesaw State University, Physics \#1202, 
       Kennesaw, GA 30144-5591,USA\\
       E-mail: \email{nkidonak@kennesaw.edu}}

\abstract{I present a calculation of higher-order radiative corrections to
charged Higgs production in association with a top quark via the process 
$bg \rightarrow tH^-$. Results for charged Higgs production at the LHC are 
presented, including the dependence of the cross section on the charged Higgs 
mass, the top quark mass, the factorization and renormalization scales, 
and $\tan \beta$. I show that the theoretical prediction for the cross section 
is significantly enhanced and is greatly stabilized when the higher-order 
corrections are included.}

\FullConference{International Europhysics Conference on High Energy Physics\\
         July 21st - 27th 2005\\
         Lisboa, Portugal}

\begin{document}

\section{Charged Higgs production via $bg \rightarrow t H^-$ }

A future observation of a charged Higgs boson at the LHC  
would be a clear sign of new physics beyond the Standard Model \cite{Higgs}.
The LHC has a good potential for discovery of a charged Higgs boson  
in association with a top quark via 
bottom-gluon fusion, $bg \rightarrow t H^-$.

The lowest-order cross section is proportional to $\alpha \alpha_s
(m_b^2\tan^2 \beta+m_t^2 \cot^2 \beta)$
where $m_b$ is the bottom quark mass, $m_t$ is the top quark mass, 
and $\tan \beta=v_2/v_1$ is
the ratio of the vacuum expectation values, $v_2$, $v_1$, 
for the two Higgs  doublets in the MSSM.
SUSY and QCD corrections to this process were calculated 
in  \cite{cHNLO}.

Threshold corrections \cite{KS,LOS} are expected to make significant 
contributions to cross sections of processes with very massive final states; 
known examples include top quark \cite{NKtop} and $W$-boson \cite {GKS} 
production, among others.
Here we calculate higher-order soft-gluon threshold corrections 
to the charged Higgs cross section through 
next-to-next-to-next-to-leading order (NNNLO) \cite{NKuni,NKchiggs}.

For the process
$b(p_b) + g(p_g) \longrightarrow t(p_t)+H^-(p_{H^-})$
we define $s=(p_b+p_g)^2$, $t=(p_b-p_t)^2$, $u=(p_g-p_t)^2$
and $s_4=s+t+u-m_t^2-{m_{H^-}}^2$.
We use the $\overline{\rm MS}$ bottom quark mass $m_b$ in the coupling
but set $m_b=0$ in the kinematics. At threshold $s_4 \rightarrow 0$. 
The soft corrections  are logarithmic terms 
of the form {$[\ln^l(s_4/{m_{H^-}}^2)/s_4]_+$}
where $l \le 2n-1$ for the order $\alpha_s^n$ corrections. 
The leading logarithms (LL) are those with 
$l=2n-1$ while the next-to-leading logarithms (NLL) are those 
with $l=2n-2$.

\section{Cross section for $bg \rightarrow t H^-$ at the LHC}

\begin{figure}
\vspace{16mm}
\begin{center}
\includegraphics[width=9.5cm,height=7.1cm]{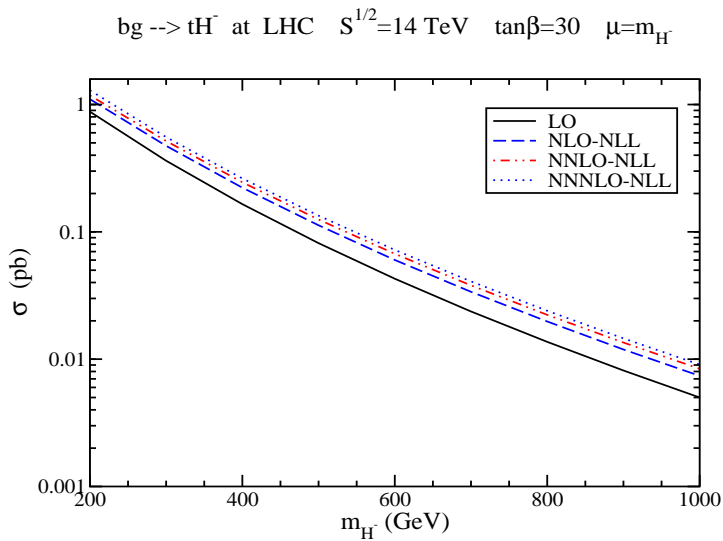}
\caption{The cross section for charged Higgs production 
via $bg \rightarrow tH^-$ at the LHC.}
\label{higgs3amhplot}
\end{center}
\end{figure}

Now we calculate the cross section for $bg \rightarrow t H^-$ at the LHC 
including NLO, NNLO, and NNNLO soft-gluon corrections at NLL accuracy.
We use the MRST2002 approximate
NNLO parton distributions functions \cite{MRST2002}
with the respective  three-loop evaluation of $\alpha_s$.

In Figure \ref{higgs3amhplot} we plot the cross section versus
charged Higgs mass for $pp$ collisions at the LHC with $\sqrt{S}=14$ TeV.
We set the factorization scale equal to the renormalization scale
and denote this common scale by $\mu$.
We show results for the LO, NLO-NLL, NNLO-NLL, and NNNLO-NLL
cross sections, all with a choice of scale $\mu=m_{H^-}$.
We use the same NNLO parton densities and couplings in all the results,
so that we can concentrate on the effects of the soft-gluon corrections.
In the calculations we choose a value $\tan \beta=30$.
It is straightforward to calculate the cross section for any other value of
$\tan \beta$, since the only dependence on $\beta$ 
is in a factor $m_b^2 \tan^2 \beta+m_t^2 \cot^2 \beta$ appearing
in the LO term.
The NLO, NNLO, and NNNLO threshold corrections are positive 
and provide a significant enhancement to the lowest-order result.
The cross sections for the related process
${\bar b} g \rightarrow {\bar t} H^+$ are exactly the same.

\begin{figure}
\vspace{16mm}
\begin{center}
\includegraphics[width=9.5cm,height=7.1cm]{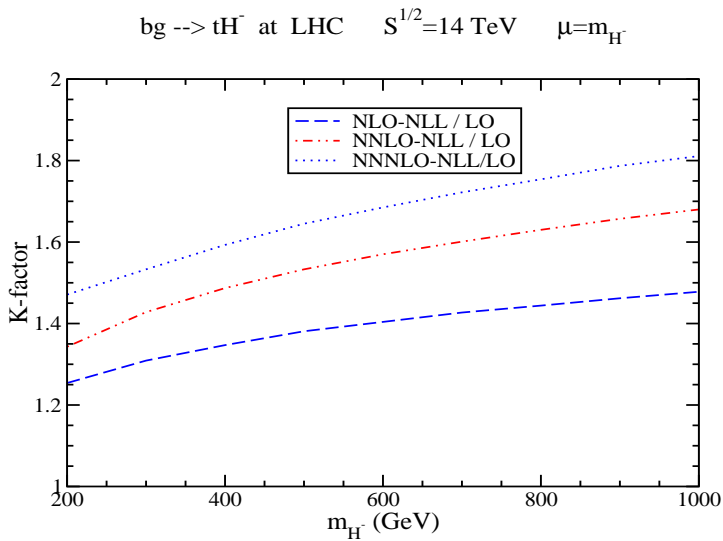}
\caption{The $K$-factors for charged Higgs production
via $bg \rightarrow tH^-$ at the LHC.}
\label{Khiggs3amhplot}
\end{center}
\end{figure}

The relative size of the corrections is better shown in Figure
\ref{Khiggs3amhplot} where we plot the $K$-factors, i.e. ratios of cross
sections at various orders. As expected
the corrections increase for higher charged Higgs masses since then
we get closer to threshold.
The NNNLO-NLL / LO curve shows that if we include all NLL threshold corrections
through NNNLO we get an enhancement over the lowest-order result of 
approximately 45\% to 80\%, depending on the charged Higgs mass. 

\begin{figure}
\vspace{16mm}
\begin{center}
\includegraphics[width=9.5cm,height=7.1cm]{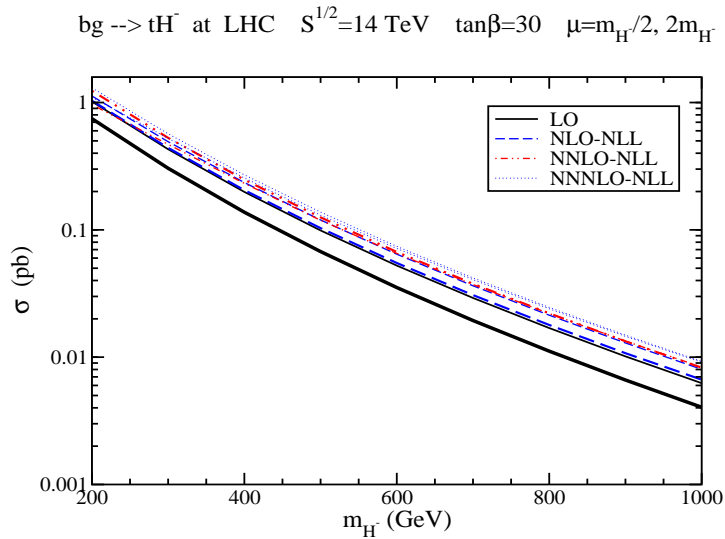}
\caption{Scale dependence of charged Higgs production
via $bg \rightarrow tH^-$ at the LHC.}
\label{higgs3amhmuplot}
\end{center}
\end{figure}

In Figure \ref{higgs3amhmuplot} the scale dependence of the charged Higgs 
cross section is shown. We see that the scale dependence diminishes 
progressively as we move from LO to NNNLO and while it is rather large at 
LO, at NNNLO it is insignificant.

Finally, we note that the cross section decreases with increasing top quark 
mass, but this dependence is not very strong.

\end{document}